# Method for including Static Correlation in Molecules

Jerry L. Whitten
Department of Chemistry
North Carolina State University
Raleigh, NC 27695 USA
email: whitten@ncsu.edu

**Abstract**

New ways to treat electron correlation in electronic structure problems are discussed in the context of many-electron theory. The present work focuses primarily on static correlation. In related work, a method for including dynamical correlation effects is described. The overlap density of two basis functions i, j and the associated density matrix, $w_{ij}$, is a signature of bond formation and can be used to define a local molecular orbital, i + j. The total electron density ρ can be written in terms of densities derived from these two-center orbitals and residual one-center terms. In the interaction of total densities $<\rho_{(1)} | r_{12}^{-1} | \rho_{(2)}>$ the self-energy terms resulting from an average field (Hartree-Fock) Hamiltonian are allowed to respond to an explicit inclusion of electron repulsion by mixing $(i + j)_1 (i + j)_2 + \lambda (i - j)_1 (i - j)_2$. The energy lowering weighted by $|w_{ij}|^2$ approximates this contribution to the correlation energy of the system. Numerical calculations for a set of 20 molecules representing different bonding environments are reported and results are compared with configuration interaction calculations using the same molecular orbital basis. Calculations on chlorin, $N_4C_{20}H_{16}$, are reported as an example of how the method could be used in an embedding treatment of a large system.

**Introduction**

The description of many-electron systems by configuration interaction (CI) is a practical way to address complex systems providing the problem can be reduced to a manageable size. There is a vast literature on ways to do this ranging from perturbation methods that generate configurations and evaluate energies efficiently to methods for partitioning large systems into localized electronic subspaces or ways to balance errors in systems that are being compared.[1-9] Relatively few configurations are required to dissociate molecules correctly or to create proper spin states. Dynamical correlation effects, particularly those associated with angular correlation, require higher spherical harmonic basis functions and this leads to a rapid increase in number of interacting configurations. In recent work, we discussed a way to include dynamical correlations by introducing a polarization of components of densities involved in Coulomb interactions.[3] The modified Coulomb interactions are used in single-determinant or configuration interaction calculations. For static correlation, one of the most important effects redistributes two-electron components to allow variable weighting of ionic and covalent contributions, giving for a symmetric bond, $\chi(1, 2) = a(1)b(2) + a(2)b(1) + \lambda[a(1)a(2) + b(2)b(1)]$, where $a$ and $b$ are functions on different nuclei. This type of correlation is critical to the description of stretched bonds and molecular dissociation. It is the simplest effect to treat by configuration interaction and the only issue is the rapid increase in number of configurations with increasing size of systems. In contrast, density functional methods do not rely on increasing the complexity of the wavefunction to achieve the requisite accuracy, but instead distribute the exchange-correlation effects over the density.

We consider in the present work a way to introduce important static correlation effects simply at the level basis functions interact to form the Coulomb repulsion of the electron density. In applications, the objective would be to treat a portion of a problem that is less important by the proposed method and reserve rigorous configuration for the region of primary importance such as a reaction site on a surface of a solid or in a large molecule.

**Method**

Consider the electron density of a system,

$$\rho(1) = \sum_p \lambda_p \varphi_p(1)\varphi_p(1) = \sum_{ij} w_{ij} f_i(1) f_j(1)$$

where $\varphi_p$ denotes a molecular orbitals expanded in terms of basis functions, $f_i$. Occupation numbers and the resulting density matrix for basis function products are denoted by $\lambda_p$ and $w_{ij}$, respectively. A signature of bond formation between basis functions $f_i$ and $f_j$ on different nuclei is the size of the overlap-density matrix product, $w_{ij} < f_i | f_j >$.

We now associate with each basis function product $f_i f_j$, a local molecular orbital, $\varphi_{ij} = (f_i + f_j)(2 + 2 < f_i | f_j >)^{-1/2}$. The density can be expressed exactly in terms of these local orbitals plus diminished contributions from $f_i f_i$,

$$\rho(1) = \sum_{ij} w'_{ij} \varphi_{ij}(1)\varphi_{ij}(1) + \sum_i w'_{ii} f_i(1) f_i(1).$$

The electron repulsion energy, $<\rho(1)|r_{12}^{-1}|\rho(2)>$, remains the same for these different decompositions of contributing interactions. In the definition of $\varphi_{ij}$, if $<f_i|f_j>$ is negative, the sign of $f_j$ is reversed.

We now proceed to alter the contributing interactions to incorporate correlation effects. Specifically, we isolate the self-energy interactions,

$$<\rho(1)|r_{12}^{-1}|\rho(2)> = \sum_{ij}|w'_{ij}|^2 <\varphi_{ij}(1)\varphi_{ij}(2)|r_{12}^{-1}|\varphi_{ij}(1)\varphi_{ij}(2)>$$
$$+ \sum_{i}|w'_{ii}|^2 <f_i(1)f_i(2)|r_{12}^{-1}|f_i(1)f_i(2)>$$
$$+ \text{ non self-energy terms}$$

and treat the product $\chi(1,2) = \varphi_{ij}(1)\varphi_{ij}(2)$ as an independent two-electron wavefunction embedded in the total electron distribution, and similarly for the products, $f_i(1)f_i(2)$. The proposed method attempts to approximate at the density and basis function level, the same correlation effects that are rigorously treated by configuration interaction at the molecular orbital level. The method departs from a rigorous argument by assuming independent local orbital excitations and requires approximations for non-orthogonality which we discuss later.

Given, $\chi(1,2) = 2^{-1/2}\det(\varphi(1)\alpha(1)\varphi(2)\beta(2))$, with the basis function pair subscript understood, and a zeroth order Hamiltonian, $H_0 = h_1 + h_2$,

$$E_0 = <\chi(1,2)|H_0|\chi(1,2)> = 2<\varphi(1)|h_1|\varphi(1)>$$

We choose $h_i$ to be the Fock operator of the self-consistent-field (SCF) solution and note that the matrix elements required are available at each SCF iteration. We now add electron repulsion explicitly, $H = H_0 + r_{12}^{-1}$, and allow the wavefunction to respond

$$\chi(1,2) = c2^{-1/2}\det(\varphi(1)\alpha(1)\varphi(2)\beta(2)) + c'2^{-1/2}\det(\varphi'(1)\alpha(1)\varphi'(2)\beta(2))$$
$$= c\chi_1(1,2) + c'\chi_2(1,2)$$

where

$$\varphi = (norm)(f_i + f_j) \quad \varphi' = (norm)(f_i - f_j) \quad <\varphi|\varphi'> = 0$$

The resulting energy lowering is defined as the correlation energy associated with the *i,j* pair bond

$$<\chi(1,2)|H_0 + r_{12}^{-1}|\chi(1,2)> - E_0$$

with a contribution to the total system weighted by $|w'_{ij}|^2$. If $f_i$ and $f_j$ are on the same nucleus, what we refer to above as a "bond" corresponds instead to a redistribution within a hybrid orbital or radial correlation of a basis function product.

We note at this point, that the idea of implementing local configuration interaction rigorously has been pursued for decades by many investigators. In early work, the idea was one of the purposes of the Edmiston - Ruedenberg localization procedure[10] and the Foster-Boys method.[11] Localization arguments have also been used by the author and co-workers to define an embedded subspace treated by configuration interaction.[12] More recent approaches can be found in Refs. 13-16. These many-electron methods are much more complex than the present simple arguments and achieve their rigor by working with an orthonormal set of orbitals and configuration interaction at the level of the total wavefunction. The question in the present work is whether correlation effects can be accounted for approximately in the framework of interactions at the basis function level plus restrictions that take into account the space already occupied in a many-electron system.

We now introduce one of the important approximations. The functions $\varphi'$ are orthogonal to the originating function $\varphi$, and to approximate the required orthogonality to the occupied orbitals of the system $\{\varphi_p\}$, we introduce a repulsive potential energy $|<\varphi'|\varphi_p>|^2 (\varepsilon' - \varepsilon_p)$, where $\varepsilon' = <\varphi'(1)|h_1|\varphi'(1)>$ and $\varepsilon_p = <\varphi_p(1)|h_1|\varphi_p(1)>$. This repulsive potential which is conceptually similar to a Phillips-Kleinman potential often used to approximate orthogonalization to core electrons is added to the energy of the correlating configuration, $\chi_2(1,2) = 2^{-1/2} \det(\varphi'(1)\alpha(1)\varphi'(2)\beta(2))$.

A second approximation concerns the residual terms $f_i f_i$ that occur in the re-expression of the density. The self-energy components of the density, $\sum_i |w'_{ii}|^2 < f_i(1)f_i(2)|r_{12}^{-1}|f_i(1)f_i(2)>$, are correlated by allowing excitations $f_i(1)f_i(2) \longrightarrow f_m(1)f_m(2)$ to available channels $f_m$. The availability of a channel is assumed to be determined by its density matrix $|w'_{mm}|^2$ and the blocking effect expressed as a reduction in exchange matrix element $(1-|w'_{mm}|^2) < f_i(1)f_i(2)|r_{12}^{-1}|f_m(1)f_m(2)>$. The correlation contribution, $<\chi(1,2)|H_0 + r_{12}^{-1}|\chi(1,2)> - E_0$, to the total system is weighted by $|w'_{ii}|^2$. Note that $w'_{ii} \leq w_{ii}$ because some of the original density was assigned to the formation of local bonds.

The basis set used in the calculations and the configuration interaction procedure used to determine the accuracy of the proposed method are described in the following sections.

**Basis set**

The basis for each atom is a near Hartree Fock set of atomic orbitals plus extra two-component s- and p-type functions consisting of the two smaller exponent components of the Hartree-Fock atomic orbital. We refer to this as a double-zeta basis. The atomic orbitals are expanded as linear combinations of Gaussian functions and the number of components in the expansions is large since the atomic orbitals are of Hartree-Fock accuracy. The basis can be described as 1s(10), 2s(5), 2p(5), 2s′(2), 2p′(2), for C, N and O, 1s(10), 2s(6), 2p(7) 2s′(2), 2p′(2), for F and 1s(4), s(1) for H where the total number of Gaussian functions for each orbital is indicated in parentheses. The larger basis sets used for $H_2$, CO, glycine and pyrazine contain additional s-, p- and d- functions.

**Configuration interaction**

All calculations are carried out for the full electrostatic Hamiltonian of the system

$$H = \sum_i^N [-\tfrac{1}{2}\nabla_i^2 + \sum_k^Q -\frac{Z_k}{r_{ik}}] + \sum_{i<j}^N r_{ij}^{-1}$$

This Hamiltonian defines the single-determinant self-consistent-field (SCF) solution and the Fock matrix elements used in the proposed method. The configuration interaction wavefunctions are constructed by multi-reference configuration interaction expansions,[7,8,12]

$$\Psi = \sum_k c_k (N!)^{-1/2} det(\chi_1^k \chi_2^k \ldots \chi_N^k) = \sum_k c_k \Phi_k$$

In all of the applications, the entire set of SCF orbitals is used to define the CI active space. Single and double excitations from the single determinant ground state wavefunction, $\Phi_r$, create a small CI expansion, $\Psi_r'$,

$$\Psi_r' = \Phi_r + \sum_{ijkl} \lambda_{ijkl} \Gamma_{ij \to kl} \Phi_r = \sum_m c_m \Phi_m$$

The configurations $\Phi_m$, are retained if the interaction with $\Phi_r$ satisfies a relatively large threshold condition

$$\frac{|\langle \Phi_m |H| \Phi_r \rangle|^2}{|E_m - E_r|} > 10^{-4} \text{ a.u.}$$

The description is then refined by generating a large CI expansion, $\Psi_r$ by single and double excitations from all important members of $\Psi_r'$ to obtain

$$\Psi_r = \Psi_r' + \sum_m \left[ \sum_{ik} \lambda_{ikm} \Gamma_{i \to k} \Phi_m + \sum_{ijkl} \lambda_{ijklm} \Gamma_{ij \to kl} \Phi_m \right]$$

where $\Phi_m$ is a member of $\Psi'_r$ with coefficient > 0.04. We refer to this expansion as a multi-reference CI. The additional configurations are generated by identifying and retaining all configurations, $\Phi_m$, that interact with $\Psi'_r$ such that

$$\frac{\left|\langle\Phi_m|\mathrm{H}|\Psi'_r\rangle\right|^2}{|E_m - E_r|} > 1\times10^{-6} \text{ a.u.}$$

For the larger molecules ~$10^5$ configurations occur in the final CI expansion, and the expansion can contain single through quadruple excitations from an initial representation of the state $\Phi_r$.

**Applications**

We consider first several molecules to obtain a preliminary understanding of the accuracy of the method for several bonding environments and to explore limitations and sensitivities to the basis set. Multi-reference CI energies are reported for comparison. Following these studies, we apply the method to a test set of 20 molecules treated only with a double-zeta s- and p-type basis. This smaller basis is more consistent with the intended use of the method.

**H₂**   Calculations on $H_2$ employ a basis set of 24 s-, p-, and d-type functions two of which are for a scaled 1s orbital. Basis functions were optimized to minimize the energy of the exact CI calculation. A calculation excluding d-type functions is included for comparison. Results are reported in Table 1 for the equilibrium internuclear distance and Fig. 1 shows a graph of the total energy vs R for the SCF and CI solutions and the current method. For both basis sets, 90% of the correlation energy is recovered. Fig. 1 shows the correct behavior on dissociation to atoms both for the CI and the proposed method, in contrast to the restricted SCF solution which dissociates incorrectly. For $H_2$, where the only uncertainty is the validity of a summation over independent excitation contributions, we consider the accuracy satisfactory.

**CO**   Two basis sets for CO are considered: a double-zeta basis and a triple-zeta plus d orbital basis. Results are reported in Table 1. The CI calculation for the small basis accounts for only about 2/3 of the correlation energy compared to the value for the larger basis. The table shows that the present method over estimates the CI correlation energy for this small basis. For the larger basis, the new method gives 78% of the correlation energy compared to the multi-reference CI calculation using the same basis. A renormalization calculation is included in the table. In this calculation, the density matrix $w_{ij}$ is set zero if a d-type function is involved and the remaining $w_{ij}$ for the s, p basis are renomalized by a common factor to conserve the total charge. The correlation energy recovered increases slightly to 80%. The renormalization calculation reveals what turns out to be a general issue. If d-type functions are added to the basis or if the original basis is uncontracted, the correlation energy from the present method deteriorates. Renormalization helps, but, for an extensively uncontracted basis, the method must be applied to groups of basis functions that resemble physical atomic orbitals. This is handled in the present work by including intact Hartree-Fock atomic orbitals in the basis.

**Glycine, $NH_2$-$CH_2$-COOH** This molecule which contains several types of bonds is treated using a double-zeta and double-zeta plus d basis. Comparing the CI results with those from the new method in Table 1 shows good agreement for the double zeta s,p basis and fairly good agreement for the basis including d-type functions; renormalization again improves the latter calculation giving a recovery of 96% of the correlation energy.

**Pyrazine, $C_4H_4N_2$** Pyrazine provides a test of the method for delocalized π-type orbitals. The double-zeta basis result is in excellent agreement with the CI calculation with an overestimate of the correlation energy of only 5%. The basis including d functions gives 81% of the correlation energy including renormalization.

It should be noted that for all of the molecules in the table, the number of bond excitations is larger than would occur for normal electron pair bonds between atoms. This means that the present method based on basis function products is correlating different regions in each electron pair bond.

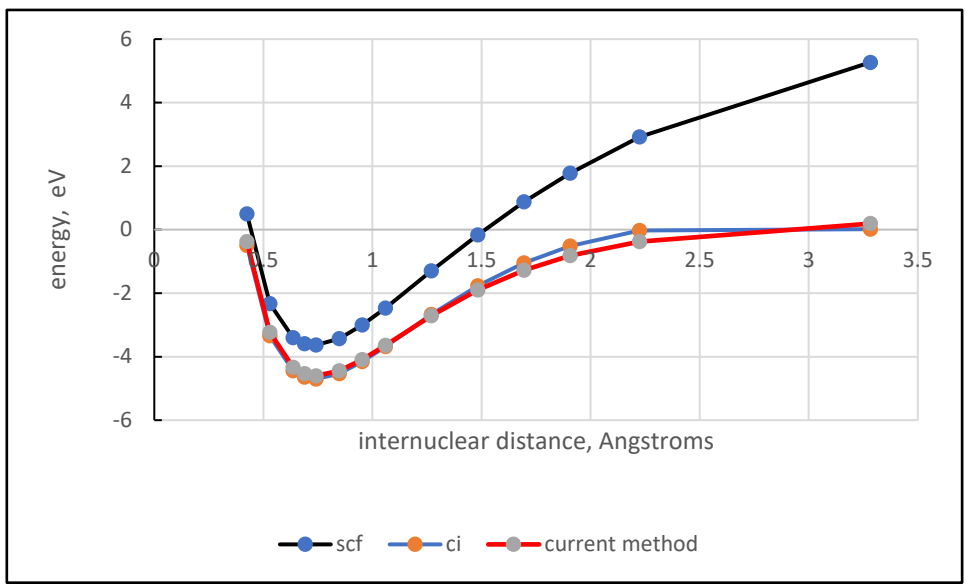

**Figure 1.** Dissociation of $H_2$. Comparison of SCF, CI and current method.

**Table 1. Comparison of correlation energies of several molecules from CI calculations and the current method.**

|  | SCF total E[a] | CI total E | current method total E | CI corr energy | current method corr energy |
|---|---|---|---|---|---|
| **H2** R=1.402 Bohr |  |  |  |  |  |
| (sss,p basis) | -1.1335 | -1.1708 | -1.1692 | -0.0373 | -0.0357 |
| (ssss,p,d basis) | -1.1336 | -1.1728 | -1.1692 | -0.0393 | -0.0356 |
| **CO** |  |  |  |  |  |
| (ss,pp basis) | -112.6983 | -112.9144 | -112.9652 | -0.2161 | -0.2669 |
| (sss,ppp,d basis) | -112.7807 | -113.0958 | -113.0256 | -0.3151 | -0.2449 |
| renormalization | -112.7807 | -113.0958 | -113.0334 | -0.3151 | -0.2527 |
| **glycine (NH2-CH2-COOH)** |  |  |  |  |  |
| (ss,pp basis) | -282.7387 | -283.1703 | -283.2054 | -0.4316 | -0.4667 |
| (ss,pp,d basis) | -282.8390 | -283.2978 | -283.2567 | -0.4588 | -0.4178 |
| renormalization | -282.8390 | -283.2978 | -283.2812 | -0.4578 | -0.4422 |
| **pyrazine (C4H4N2)** |  |  |  |  |  |
| (ss,pp basis) | -262.5800 | -263.0770 | -263.0873 | -0.4970 | -0.5073 |
| (ss,pp,d basis) | -262.6764 | -263.2787 | -263.1253 | -0.6024 | -0.4489 |
| renormalization | -262.6764 | -263.2787 | -263.1639 | -0.6024 | -0.4875 |

| calculation details | number of basis orbitals | CI dets | bond excitations | current method bond contribution[a] | 1-ctr contribution[a] |
|---|---|---|---|---|---|
| **H2** |  |  |  |  |  |
| (sss,p basis) | 12 | 21 | 7 | -0.0336 | -0.0021 |
| (ssss,p,d basis) | 24 | 100 | 7 | -0.0334 | -0.0022 |
| **CO** |  |  |  |  |  |
| (ss,pp basis) | 18 | 9736 | 21 | -0.1174 | -0.1495 |
| (sss,ppp,d basis) | 36 | 49009 | 41 | -0.0844 | -0.1605 |
| renormalization | 36 | 49009 | 34 | -0.0869 | -0.1658 |
| **glycine** |  |  |  |  |  |
| (ss,pp basis) | 55 | 89711 | 315 | -0.2855 | -0.1812 |
| (ss,pp,d basis) | 85 | 158031 | 413 | -0.2341 | -0.1837 |
| renormalization | 85 | 158031 | 297 | -0.2452 | -0.1970 |
| **pyrazine** |  |  |  |  |  |
| (ss,pp basis) | 62 | 93140 | 351 | -0.3583 | -0.1490 |
| (ss,pp,d basis) | 92 | 118394 | 477 | -0.2966 | -0.1523 |
| renormalization | 92 | 118394 | 376 | -0.3218 | -0.1657 |

[a]Energies are in hartrees.

**Other molecules**

We consider next a set of seventeen molecules containing C, N, O, F and H chosen to represent different types of bonds: single and multiple bonds, strained bonds, aromatic and non-aromatic systems. It is not a large data set, but one that should be adequate to identify failures of the method or delineate limitations.  In this series of calculations, only a double-zeta basis is used, and the CI values used for comparison are from single and double excitations from the ground state determinant as defined above. Extrapolation procedures are used to provide an estimate of the single and double excitation limit.[7,8] Results are tabulated in Table 2. Values from multi-reference CI calculations are also included in the table.

An examination of the results in Table 2 shows for most molecules fairly good agreement between the correlation energy calculated by CI and the proposed method using the same basis. For all molecules in the table, at least 80% of the correlation energy calculated by CI is recovered.  For several molecules the correlation energy compared to the single and double excitation value is overestimated. The error is less compared to the multi-reference CI, but, there is no justification for assuming this simple method should recover subtle correlation effects associated with multi-reference calculations.   We conclude that static correlation effects are fairly well approximated for this set of molecules, but that the errors are not uniform.

**Table 2.  Comparison of correlation energies from the present method with CI values.**

|  | **NC4H4N** | **NC5H5** | **C6H6** | **HCCH** |
|---|---|---|---|---|
| SCF total energy (hartrees) | -262.5800 | -246.6176 | -230.6485 | -76.8089 |
| correlation energy (hartrees) | | | | |
| present method | -0.5073 | -0.4964 | -0.4983 | -0.1752 |
| 1,2 excitation-CI | -0.4867 | -0.4701 | -0.4631 | -0.1865 |
| multi-ref CI | -0.5178 | -0.4966 | -0.4866 | -0.2031 |
| % error[a] | -4.2 | -5.6 | -7.6 | 6.1 |
|  | **CH4** | **C2H4** | **NC4H4** | **NC4H5** |
| SCF total energy (hartrees) | -40.1874 | -78.0194 | -208.0959 | -208.7468 |
| correlation energy (hartrees) | | | | |
| present method | -0.1156 | -0.1566 | -0.3663 | -0.3978 |
| 1,2 excitation-CI | -0.1142 | -0.1931 | -0.4102 | -0.4184 |
| multi-ref CI | -0.1164 | -0.2055 | -0.4270 | -0.4272 |
| % error[a] | -1.2 | 18.9 | 10.7 | 4.9 |

|  | NH2-CH2-COOH | H2CO | H2O | C6H5-COOH |
|---|---|---|---|---|
| SCF total energy (hartrees) | -282.7387 | -113.8287 | -76.0079 | -418.1783 |
| correlation energy (hartrees) |  |  |  |  |
| present method | -0.4667 | -0.1758 | -0.1034 | -0.7944 |
| 1,2 excitation-CI | -0.4633 | -0.2147 | -0.1267 | -0.6667 |
| multi-ref CI | -0.4729 | -0.2286 | -0.1280 | -0.6890 |
| % error[a] | -0.7 | 18.1 | 18.4 | -19.1 |

|  | C5H5-COOH | C6H5-F | FHCO | C2F2H2 |
|---|---|---|---|---|
| SCF total energy (hartrees) | -380.2957 | -329.5020 | -212.6782 | -275.6546 |
| correlation energy (hartrees) |  |  |  |  |
| present method | -0.7110 | -0.6109 | -0.2618 | -0.3290 |
| 1,2 excitation-CI | -0.6225 | -0.5493 | -0.3253 | -0.4087 |
| multi-ref CI | -0.6523 | -0.5788 | -0.3421 | -0.4292 |
| % error[a] | -14.2 | -11.2 | 19.5 | 19.5 |

|  | C6H5-NH2 |
|---|---|
| SCF total energy (hartrees) | -285.6597 |
| correlation energy (hartrees) |  |
| present method | -0.5761 |
| 1,2 excitation-CI | -0.5306 |
| multi-ref CI | -0.5575 |
| % error[a] | -8.6 |

[a]Errors are relative to the 1,2-excitation CI values.

**Embedding application – chlorin**

In this section, we illustrate the intended used of the present method to describe a portion of a problem that is assumed to be of secondary importance to an embedded subspace that is to be treated by a more accurate method. We use chlorin, depicted in Fig. 2, as an example and define a subspace consisting of the central region of the molecule, four nitrogen and two hydrogen atoms, plus the entire delocalized π-system. An embedded subspace encompassing this region is defined by transforming the delocalized molecular orbitals from the SCF solution to maximize their interaction with basis functions on the N, H nuclei and the p-π orbitals of the molecule. This can be accomplished either by maximizing the exchange interaction or the overlap as described in Ref. 12 The result of this localizing transformation is a set of 30 localized molecular orbitals, (60 electrons), and a complementary set of 52 orbitals, (104 electrons), that defines the region of secondary interest. Occupied and virtual orbitals are transformed separately. The transformation is unitary and thus leaves the total wavefunction invariant. The plan is to use configuration interaction to describe the localized subspace of 60 electrons and the present method to describe the remainder of the system containing 52 molecular orbitals, (104 electrons). The density matrix that drives the present method now comes from these 52 occupied molecular orbitals, but the orthogonality projector must make use of the entire set of occupied molecular orbitals. It is straightforward to use the present method to describe this 52 molecular orbital subspace. However, it is more difficult to obtain a configuration interaction value for comparison. To do this, the 24 1s orbitals are removed from the CI subspace and a new set of virtual orbitals is obtained by exchange maximization with the remaining 52-24=28 molecular orbitals.[12] This is a useful way to reduce the CI problem to a manageable size. The results and comparison checks are summarized below (energies in hartrees).

    60 electron primary system:        -0.4439  (correlation energy by explicit CI)
    104 electron secondary system:    -0.5410  (correlation energy by present method)
    104 electron secondary system:    -0.5638  (correlation energy check by CI)

In practice, if there were confidence in the proposed method, the latter CI calculation would not be performed. As a further numerical check, if the present method is used for the primary system the correlation energy calculated is -0.4787 instead of -0.4439 from CI. It is gratifying that for such a complicated subspace the error in the present method is only a few percent, comparable to the values reported in the other studies.

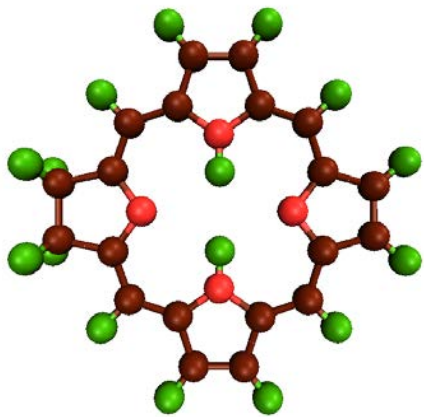

**Figure 2.** The chlorin molecule and its extensively delocalized π-electron system

**Conclusion**

      The proposed method appears to show promise as an approximate method particularly for smaller, double-zeta basis sets. The primary objective of the method is to introduce static effects associated with bond dissociation and this objective has been accomplished for the molecules studied. There is an ambiguity in the sense that some dynamical correlation effects are included in both methods, in the CI when excitations to higher spherical harmonic molecular orbitals occur such as s → p excitations and p → d excitations in the extended basis calculations. If we

want to use the present method and to include dynamical correlation by the method discussed in Ref. 3, it would be necessary to adjust one of the methods so that effects are not double counted. This should be investigated in future work. For extensively uncontracted basis sets and those that include higher spherical harmonic functions, renormalization is useful but it remains necessary to group functions prior to application of the method to replicate principal atomic orbital contributions to bonds. As demonstrated for chlorin, the method is simple to implement in an embedding framework to describe a portion of a problem that is of secondary importance and to use a rigorous CI method for the subspace of primary importance.